\documentclass[10pt, journal]{IEEEtran}
\usepackage[utf8]{inputenc}
\usepackage{amsmath, amssymb}
\usepackage{color}
\usepackage{xcolor}
\usepackage{comment}
\usepackage{url}
\usepackage{multirow}
\usepackage{stmaryrd}
\usepackage{xspace}
	\usepackage{adjustbox}

\usepackage[
n,
operators,
advantage,
sets,
adversary,
landau,
probability,
notions,	
logic,
ff,
mm,
primitives,
events,
complexity,
asymptotics,
keys]{cryptocode}

\usepackage{tcolorbox}

\tcbuselibrary{skins,breakable}
\newtcolorbox{functionality}[2][]{%
  enhanced,
  title        = {#2},
  attach boxed title to top left={xshift=+3mm,yshift*=-3mm},
  breakable    = true,
  colback      = blue!5,
  colframe     = blue!35!black,
  fonttitle    = \bfseries,
  colbacktitle = blue!15!white,
  coltitle     = black,
  #1
}

\newcommand{\sharetwo}[1]{\ensuremath{\llbracket{#1}\rrbracket_{_2}}\xspace}
\newcommand{\shareq}[1]{\ensuremath{\llbracket{#1}\rrbracket_{_q}}\xspace}
\def\getsr{\stackrel{{\scriptscriptstyle\$}}{\leftarrow}}
\newcommand{\fti}[1]{\ensuremath{\mathcal{F}^{\mathcal{D}_{#1}}_{\mathsf{TI}}}\xspace}
\newcommand{\fconv}{\ensuremath{\mathcal{F}_{\mathsf{2toQ}}}\xspace}
\newcommand{\feq}{\ensuremath{\mathcal{F}_{\mathsf{EQ}}}\xspace}
\newcommand{\ffe}{\ensuremath{\mathcal{F}_{\mathsf{FE}}}\xspace}
\newcommand{\fbtx}{\ensuremath{\mathcal{F}_{\mathsf{BTX}}}\xspace}
\newcommand{\fppnbc}{\ensuremath{\mathcal{F}_{\mathsf{PPNBC}}}\xspace}
\newcommand{\peq}{\ensuremath{\pi_{\mathsf{EQ}}}\xspace}
\newcommand{\pfe}{\ensuremath{\pi_{\mathsf{FE}}}\xspace}
\newcommand{\pconv}{\ensuremath{\pi_{\mathsf{2toQ}}}\xspace}

\newcommand{\fgeq}{\ensuremath{\mathcal{F}_{\mathsf{GEQ}}}\xspace}
\newcommand{\fmmul}{\ensuremath{\mathcal{F}_{\mathsf{DMM}}}\xspace}
\newcommand{\pdecompopt}{\ensuremath{\pi_{\mathsf{decompOPT}}}\xspace}
\newcommand{\s}{\ensuremath{\mathcal{S}}\xspace}
\newcommand{\env}{\ensuremath{\mathcal{Z}}\xspace}
\newcommand{\F}{\ensuremath{\mathcal{F}}\xspace}
\newcommand{\Zqm}[2]{\ensuremath{\mathbb{Z}_q^{#1\times #2}}\xspace}

\newcommand{\change}[1]{#1}

\begin{document}
\title{Fast Privacy-Preserving Text Classification based on Secure Multiparty Computation}

\author{Amanda Resende,
        Davis Railsback,
        Rafael Dowsley,
        Anderson C. A. Nascimento
        and~Diego F. Aranha
        
\thanks{A. Resende is with the {Institute of Computing, University of Campinas, Campinas, Brazil}. E-mail: {amanda.resende@ic.unicamp.br}.}%
\thanks{D. Railsback and A. C. A. Nascimento are with the School of Engineering and Technology, University of Washington Tacoma, WA 98402, USA. E-mails: \{drail, andclay\}@uw.edu.}%
\thanks{R. Dowsley is with the Faculty of Information Technology, Monash University, Australia. E-mail: rafael.dowsley@monash.edu.}%

\thanks{D. F. Aranha is with the Department of Computer Science, Aarhus University, Aarhus, Denmark. E-mail: dfaranha@cs.au.dk.}%

}

\maketitle

\begin{abstract}


We propose a privacy-preserving Naive Bayes classifier and apply it to the problem of private text classification. In this setting, a party (Alice) holds a text message, while another party (Bob) holds a classifier. At the end of the protocol, Alice will only learn the result of the classifier applied to her text input and Bob learns nothing. Our solution is based on Secure Multiparty Computation (SMC). Our Rust implementation provides a fast and secure solution for the classification of unstructured text. Applying our solution to the case of spam detection (the solution is generic, and can be used in any other scenario in which the Naive Bayes classifier can be employed), we can classify an SMS as spam or ham in less than 340\,ms in the case where the dictionary size of Bob's model includes all words ($n = 5200$) and Alice's SMS has at most $m = 160$ unigrams. In the case with $n = 369$ and $m = 8$ (the average of a spam SMS in the database), our solution takes only 21\,ms.

\end{abstract}


\section{Introduction}

Classification is a supervised learning technique in Machine Learning (ML) that has the goal of constructing a classifier given a set of training data with class labels. Decision Tree, Naive Bayes, Random Forest, Logistic Regression and Support Vector Machines (SVM) are some examples of classification algorithms. These algorithms can be used to solve many problems, such as: classifying an email/Short Message Service (SMS) as spam or ham (not spam)~\cite{Almeida11}; diagnosis of a medical condition (disease versus no disease)~\cite{Voets18}; hate speech detection~\cite{NeurIPS2019}; face classification~\cite{Kumar11}; fingerprinting identification~\cite{Cappelli10}; and image categorization~\cite{Deng10}. For the first three examples above, classification is binary, where there are only two class labels (yes or no); while the last three are multi-class, that is, there are more than two classes.

We consider the scenario in which there are two parties: one possesses the private data to be classified and the other party holds a private model used to classify such data. In such a scenario, the party holding the data (Alice) is interested in obtaining the classification result of such data against a model held by a second party (Bob) so that, at the end of the classification protocol, Alice knows solely the input data and the classification result, and Bob knows nothing beyond the model itself. This scenario is a very relevant one. There are many situations where a data owner is not comfortable sharing a piece of data that needs classification (think of psychological or health related data). Also, a machine learning model holder might not want to/cannot reveal the model in the clear for intellectual property issues or because the model reveals information about the data set used to train it. Thus, both parties have proper incentives to participate in a protocol providing the joint functionality of private classification. 


Due to these concerns, mechanisms such as Secure Multiparty Computation (MPC)~\cite{CDN2015}, Differential Privacy (DP) and Homomorphic Encryption (HE) can be used to build privacy-preserving solutions. MPC allows two or more parties to jointly compute a function over their private inputs without revealing any information to the other party, whereas HE is an encryption scheme that allows performing computations on encrypted data without having to decrypt it. And, DP is a technique that adds random noise to queries, to prevent an adversary from learning information about any particular individual in the data set.

Our main goal is to propose protocols for privacy-preserving text classification. By carefully selecting \change{cryptographic} engineering optimizations, we improve upon previous results by Reich \textit{et al.}~\cite{NeurIPS2019} by over one order of magnitude achieving, to the best of our knowledge, the fastest text-classification results in the available literature (21ms for an average sample of our data set). More specifically, we propose a privacy-preserving Naive Bayes classification (PPNBC) based on MPC where given a trained model we classify/predict an example without revealing any additional information to the parties other than the classification result, which can be revealed to one specified party or both parties. We then apply our solution to a text classification problem: classifying SMSes as spam or ham.

\subsection{Application to Private SMS Spam Detection}

SMS is one of the most used telecommunication service in the world. It allows mobile phone users to send and receive a short text (which has 160 7-bit characters maximum). Due to advantages such as reliability (since the message reaches the mobile phone user), low cost to send an SMS (especially if bought in bulk), the possibility of personalizing, and immediate delivery, SMS is a widely used communication medium for commercial purposes, and mobile phone users are flooded with unsolicited advertising.

SMSes are also used in scams, where someone tries to steal personal information, such as credit card details, bank account information, or social security numbers. Usually, the scammer sends an SMS with a link that invites a person to verify his/her account details, make a payment, or that claims that he/she has earned some amount of money and needs to use the link to confirm. In all cases, such SMSes can be classified as spam.

Machine learning classifiers can be used to detect whether an SMS is a spam or not (ham). During the training phase, these algorithms learn a model from a data set of labeled examples, and later
on, are used during the classification/prediction phase to classify unseen SMSes. In a Naive Bayes classifier, the model is based on the frequency that each word occurs in the training data set. In the classification phase, based on these frequencies, the model predicts whether an unseen SMS is spam or not.

A concern with this approach is related to Alice's privacy since she needs to make her SMSes available to the spam filtering service provider, Bob, which owns the model. SMSes may contain sensitive information that the user would not like to share with the service provider. Besides, the service provider also does not want to reveal what parameters the model uses (in Naive Bayes, the words and its frequencies) to spammers and concurrent service providers. Our privacy-preserving Naive Bayes classification based on MPC, provides an extremely fast secure solution for both parties to classify SMSes as spam or ham without leaking any additional information while maintaining essentially the same accuracy as the original algorithm performed in the clear. While our experimental treatment is focused on SMS messages, the same approach can be naturally generalized to classify short messages received over Twitter or instant messengers such as WhatsApp or Signal.

\subsection{Our Contributions}	

These are the main contributions of this work: 

\begin{itemize}

	\item \textbf{A privacy-preserving Naive Bayes classification (PPNBC) protocol}: We propose the first  privacy-preserving Naive Bayes classifier with private feature extraction. Previous works assumed the features to be publicly known. It is based on secret sharing techniques from MPC. In our solution, given a trained model it is possible to  classify/predict an example without revealing any additional information to the parties other than the classification result, which can be revealed to one or both parties. \change{We prove that our solution is secure using the Universal Composability (UC) framework~\cite{Canetti01Uni}, the gold standard framework for analyzing the security of cryptographic protocols, thus proving that it enjoys very strong security guarantees, and can be arbitrary composed as well as used in realistic scenarios such as the Internet without compromising the security of the solution.
	}  
	
	\item \textbf{An efficient and optimized software implementation of the protocol:} The proposed protocol is implemented in Rust using an up-to-date version of the RustLynx framework available at \url{https://bitbucket.org/uwtppml/rustlynx/src/master/}. 
	 
	\item \textbf{Experimental results for the case of SMS classification as spam/ham}: The proposed protocol is evaluated in a use case for SMS spam detection, using a data set widely used in the literature. However, it is important to note that the solution is generic, and can be used in any other scenario in which the Naive Bayes classifier can be employed. 

\end{itemize}

While the necessary building blocks already exist in the literature, the main novelty of our work is putting these building blocks together, optimizing their implementations \change{using cryptographic engineering techniques}
and obtaining the fastest protocol for private text classification to date. 

\subsection{Organization} This paper is organized as follows. In Section 2, we define the notation used during this work, describe the necessary cryptographic building blocks and briefly introduce the Naive Bayes classifier. In Sections 3 and 4, we describe our privacy-preserving Naive Bayes classification protocol and present its security analysis. In Section 5, we describe the experimental results, from the training phase until the classification using our PPBNC, as well as the cryptographic engineering techniques used in our implementation. Finally, in Sections 6 and 7, we present the related works and the conclusions.

\section{Preliminaries}

As in most existing works on privacy-preserving machine learning based on MPC, we consider an honest-but-curious adversary (also known as semi-honest adversary). In this model, each party follows the protocol specifications \change{but, using his view of the protocol execution, may try to learn additional information other than his input and specified output.}

\subsection{Secure Computation Based on Additive Secret Sharing}\label{sec:secsharing}

Our solution is based on additive secret sharing over a ring $\mathbb{Z}_q = \{0, 1 , 2, \cdots, q-1\}$. A value $x$ is secret shared between Alice and Bob over $\mathbb{Z}_q$ by picking $x_A, x_B \in \mathbb{Z}_q$ uniformly at random with the constraint that $x = x_A + x_B$ mod $q$.  Alice receives the share $x_A$ while Bob receives the share $x_B$. We denote this pair of shares by \shareq{x}. A secret shared value $x$ can be open towards one party by disclosing both shares to that party. Let $\shareq{x}$, $\shareq{y}$ be secret shared values and $c$ be a constant. Alice and Bob can perform locally and straightforwardly the following operations:

\begin{itemize}
	
	\item Addition ($z = x + y$): Each party locally adds its local shares of $x$ and $y$ modulo $q$ in order to obtain a share of $z$. This operation will be denoted by $\shareq{z} \leftarrow \shareq{x} + \shareq{y}$.
	
	\item Subtraction ($z = x - y$): Each party locally subtracts its local share of $x$ and $y$ modulo $q$ in order to obtain a share of $z$. This operation will be denoted by $\shareq{z} \leftarrow \shareq{x} - \shareq{y}$.
	
	\item Multiplication by a constant ($z = cx$): Each party multiplies its local share of $x$ by $c$ modulo $q$ to obtain a share of $z$. This operation will be denoted by $\shareq{z} \leftarrow c \shareq{x}$.
	
	\item Addition of a constant ($z = x + c$): Alice adds $c$ to her share $x_A$ of $x$ to obtain $z_A$, while Bob sets $z_B$ = $x_B$, keeping his original share. This will be denoted by $\shareq{z} \leftarrow \shareq{x} +c$.
	
\end{itemize}

Unlike the above operations, secure multiplication of secret shared values (i.e., $z = xy$) cannot be done locally and requires interaction between Alice and Bob. To compute this operation highly efficiently, we use Beaver's multiplication triple technique~\cite{Beaver97}, which  consumes a multiplication triple \change{-- i.e., $(\shareq{u},\shareq{v},\shareq{w})$ such that $u$ and $v$ are chosen uniformly at random and $w=uv$ --} in order to compute the multiplication of $\shareq{x}$ and $\shareq{y}$ without leaking any information. We use a trusted initializer (TI) to generate multiplication triples and secret share them to Alice and Bob. In the trusted initializer model, the TI can pre-distribute correlated randomness to Alice and Bob during a setup phase, which is run before the protocol execution (possibly long before Alice and Bob get to know their inputs). The TI is not involved in any other part of the protocol execution and does not get to know the parties inputs and outputs.\footnote{It is a well-known fact that UC-secure MPC needs a setup assumption \cite{C:CanFis01,STOC:CLOS02}. A TI is one of the setup assumptions that allows obtaining UC-secure MPC. Other setup assumptions that enable UC-secure MPC include: a common reference string \cite{C:CanFis01,STOC:CLOS02,C:PeiVaiWat08}, the availability of a public-key infrastructure \cite{FOCS:BCNP04}, signature cards \cite{HofMulUhr05}, tamper-proof hardware \cite{EC:Katz07,TCC:DotKraMul11,ICITS:DowMulNil15} or noisy channels between the parties \cite{SBSEG:DMN08,JIT:DGMN13}, and the random oracle model \cite{TCC:HofMul04,EPRINT:BDDMN17b,CANS:DavDow20}.
}  This model was used in many previous works, e.g., \cite{r99,dowsley2010two,IEICE:DMOHIN11,ishai2013power,IJIS:TNDMIHO15,david2015efficient,IEEEIFS:DDGM+16,fritchman2018,IEEENSRE:ADMW+19}. If a TI is not desirable or unavailable, Alice and Bob can securely simulate a TI at the cost of introducing computational assumptions in the protocol~\cite{IEEETDSC:CDHK+19}. The TI is modeled by the ideal functionality $\fti{}$. 
In addition to the multiplication triples, the TI also generates random values in $\mathbb{Z}_q$ and delivers them to Alice so that she can use them to secret share her inputs. If Alice wants to secret share an input $x$, she picks an unused random value $r$ (note that Bob does not know $r$), and sends $c=x-r$ to Bob. 
Her share $x_A$ of $x$ is then set to $x_A=r$, while Bob's share $x_B$ is set to $x_B=c$. The secret sharing of Bob's inputs is done similarly using random values that the TI only delivers to him.

\begin{functionality}{\textbf{Functionality} $\fti{}$}
{
$\fti{}$ is parametrized by an algorithm $\mathcal{D}$. Upon initialization run $(D_A, D_B) \getsr \mathcal{D}$ and deliver $D_A$ to Alice and $D_B$ to Bob.
}
\end{functionality}

The following straightforward extension of Beaver's idea performs the UC-secure multiplication of secret shared matrices $X \in \mathbb{Z}_q^{i \times j}$ and $Y \in \mathbb{Z}_q^{j \times k}$~\cite{Dowsley16,IEEETDSC:CDHK+19}. The protocol will be denoted by $\pi_{\mathsf{DMM}}$ and works as follows:

\begin{enumerate}
	
	\item The TI chooses uniformly random $U$ and $V$ in $\mathbb{Z}_q^{i \times j}$ and $\mathbb{Z}_q^{j \times k}$, respectively, computes $W = UV$ and pre-distributes secret sharings $\shareq{U}, \shareq{V}, \shareq{W}$ (the secret sharings are done element-wise) to Alice and Bob.
	
	\item Alice and Bob locally compute $\shareq{D} \leftarrow \shareq{X} - \shareq{U}$ and $\shareq{E} \leftarrow \shareq{Y} - \shareq{V}$, and then open $D$ and $E$.
	
	\item Alice and Bob locally compute $\shareq{Z} \leftarrow \shareq{W} + E\shareq{U} + D \shareq{V} + DE$.
\end{enumerate}

\subsection{Fixed Point Arithmetic}\label{section:fixed_point_arithmetic}

Many real-world applications of MPC require representing and operating on continuous data. This poses a challenge because the security of additive secret sharing depends on the fact that shares are uniformly random -- a concept that only exists for samples of finite sets. For compatibility with MPC, continuous values need to be represented within a range of possible values. We use the mapping
\begin{equation*}
Q(x) = 
\begin{cases} 
      2^\lambda - \left \lfloor{ 2^a \cdot |x|  }\right \rfloor & \mbox{if\ } x < 0 \\
      \left \lfloor{ 2^a \cdot x  }\right \rfloor & \mbox{if\ } x \geq 0
\end{cases}
\end{equation*}
to represent real numbers in $\mathbb{Z}_q$, where $q = 2^\lambda$, as fixed-point precision two's complement values. The parameter $a$ is the \textit{fractional accuracy} -- the number of bits used to represent negative powers of 2. This mapping preserves addition in $\mathbb{Z}_q$ straightforwardly, but a multiplication of two fixed-point values results in a fixed point value with $2a$ fractional bits. To maintain the expected representation in $\mathbb{Z}_q$, all products need to be truncated by $a$ bit positions, requiring an additional MPC protocol \cite{mohassel2017secureml}. In this paper, fixed-point values are only added together and multiplied by 0 or 1, so a truncation protocol is not needed for our purposes.

\subsection{Cryptographic Building Blocks}\label{building blocks}

Next, we present the cryptographic building blocks that are used in our PPNBC solution. 

\noindent\textbf{Secure Equality Test:} To perform a secure equality test, we use a straightforward folklore protocol $\peq$. As input, Alice and Bob have bitwise secret sharings in $\mathbb{Z}_2$ of the bitstrings $X = \{x_\ell, \ldots, x_1\}$ and $Y = \{y_\ell, \ldots, y_1 \}$. The protocol generates as output a secret sharing of 1 if $X = Y$ and a secret sharing of 0 otherwise. The protocol $\peq$ works as follows:
	
\begin{enumerate}
		
	\item For $ i = 1, \ldots, \ell$, Alice and Bob locally compute $\sharetwo{r_i} \gets \sharetwo{x_i} + \sharetwo{y_i} + 1$. 
		
	\item Alice and Bob use secure multiplication ($\pi_{\mathsf{DMM}}$) to compute a secret sharing of $z = r_1 \cdot \ldots \cdot r_\ell$. They output the secret sharing $\sharetwo{z}$. (Note that if $x = y$, then $r_i = 1$ in all $i$ positions, thus $z =1$; otherwise some $r_i = 0$ and so $z = 0$).
\end{enumerate}

By performing the multiplications to compute $z$ in a binary tree style with the values $r_1, \ldots, r_\ell$ in the $\ell$ leaves, 
the protocol $\peq$ requires $\lceil \log(\ell) \rceil$ rounds of communication and a total of $4 (\ell -1)$ bits of data transfer. For batched inputs $\{X_1, ..., X_k \}$, $\{Y_1, ..., Y_k \}$, the number of communication rounds remains the same and the data transfer per round is scaled by $k$.

\noindent\textbf{Secure Feature Extraction:} To perform the feature extraction in a privacy-preserving way, we use the protocol $\pfe$ from Reich et al.~\cite{NeurIPS2019}. Alice has as input the set $A=\{a_1, \cdots, a_m\}$ of unigrams occurring in her message and Bob has as input the set $B=\{b_1, \cdots, b_n \}$ of unigrams that occur in his ML model. The elements of both sets are represented as bitstrings of size $\ell$.
The purpose of the protocol is to extract which words from Alice's message appear in Bob's set. Thus, at the end of the protocol, Alice and Bob have secret shares of a binary feature vector $Y$ which represents what words in Bob’s set appear in Alice’s message. The binary feature vector $Y = \{y_1, \cdots, y_n \}$ is defined as:
\[y_j = \left \{ \begin{matrix} 1, & \mbox{if }b_i\in \mbox{A} \\ 0, & \mbox{otherwise} \end{matrix} \right.\]
	
The protocol $\pfe$ works as follows:
	
\begin{enumerate}
		
	\item Alice secret shares $a_p$ with Bob  for $p=1, \cdots, m$, while Bob secret shares $b_i$ with Alice for $i=1, \cdots, n$. Both use bitwise secret sharings in $\mathbb{Z}_2$. To secret share their input $a_p$ and $b_i$, Alice and Bob use the method described in Section \ref{sec:secsharing}.
		
	\item For each $a_p$ and each $b_i$, they execute the secure equality protocol $\peq$, which outputs a secret sharing of
	\[y'_{ip} = \left \{ \begin{matrix} 1, & \mbox{if }a_p \mbox{ = } b_i \\ 0, & \mbox{otherwise} \end{matrix} \right.\]

	\item Alice and Bob locally compute the secret share $\sharetwo{y_i} \leftarrow \sum_{p=1}^{m} \sharetwo{y'_{ip}}$
		
\end{enumerate}

The protocol requires $\lceil \log(\ell) \rceil +1$ rounds of communication, $4 (\ell -1)$ bits of data transfer for each call of the $\peq$, and 1 bit for each input bit that is secret shared. $\pfe$ requires $ m \cdot n $ equality tests: the number of communication rounds remains the same as a single execution, as all the tests can be done in parallel; the data transfer however is scaled by $m \cdot n$. The total data transfer of $\pfe$ is therefore $4 \cdot m \cdot n \cdot (\ell -1) + m + n$ bits.

\noindent\textbf{Secure Conversion:} To perform a secure conversion from a secret sharing in $\mathbb{Z}_2$ to a secret sharing in $\mathbb{Z}_q$, we use the secure conversion protocol $\pconv$ presented by Reich \textit{et al.}~\cite{NeurIPS2019}. Alice and Bob have as input a secret sharing $\sharetwo{x}$ and without learning any information about $x$, they  must get a secret sharing $\shareq{x}$. The protocol $\pconv$ works as follows:
	
\begin{enumerate}
		
	\item For the input $\sharetwo{x}$, let $x_A \in \{0, 1\}$ denote Alice's share of $x$ and $x_B \in \{0, 1\}$ denote Bob's share.
	
	\item Define $\shareq{x_A}$ as the shares $(x_A,0)$ and $\shareq{x_B}$ as the shares $(0,x_B)$.
	
	\item Alice and Bob compute $\shareq{y} \gets \shareq{x_A}\shareq{x_B}$.
		
	\item They output $\shareq{z} \gets \shareq{x_A} + \shareq{x_B}- 2\shareq{y}$.
		
\end{enumerate}

The protocol $\pconv$ requires $1$ round of communication and a total of $4\lambda$ bits of data transfer, where $\lambda$ is the bit length of $q$. For batched inputs $\{ x_1, ..., x_k \}$, the number of communication rounds remains the same and the data transfer is scaled by $k$.

\noindent\textbf{Secure Bit Extraction:} The secure bit extraction protocol $\pi_{\mathsf{{BTX}}}$ takes a secret value $\shareq{x}$ and a publicly known bit position $\alpha$ and returns a $\mathbb{Z}_2$-sharing of the $\alpha$-th bit of $x$, $\sharetwo{(x >> (\alpha-1)) \; \land \; 1}$. The protocol is based on a reduction of the protocol for full bit decomposition modeled after a matrix representation of the carry look-ahead adder circuit that was presented in \cite{Cock20}. $\pi_{\mathsf{{BTX}}}$ \cite{unpublishedDT}
works as follows:

\begin{enumerate}
	\item For the secret shared value $\shareq{x}=(x_A,x_B)$, Alice and Bob locally create bitwise sharings of the \textit{propagate signal}
	\[ p^{(\alpha)} \gets x_A^{(\alpha)} \oplus x_B^{(\alpha)}, \; \ldots, \; p^{(1)} \gets x_A^{(1)} \oplus x_B^{(1)}, \]
	where $x_A^{(i)}/x_B^{(i)}$ indicates the $i$-th bit of $x_A/x_B$.
	
	\item Alice and Bob use the secure multiplication of secret shared values to jointly compute the \textit{generate signal}
	\[g^{(\alpha)} \gets x_A^{(\alpha)} \land x_B^{(\alpha)}, \; \ldots, \; g^{(1)} \gets x_A^{(1)} \land x_B^{(1)}. \]
	
	\item Alice and Bob jointly compute the $(\alpha-1)$-th \textit{carry bit} 
	$\sharetwo{c^{(\alpha-1)}}$ as the upper right entry of $$\prod_{1 \leq i \leq \alpha-1} \begin{bmatrix}
	\sharetwo{p^{(i)}} & \sharetwo{g^{(i)}} \\
	0 & 1 \\
	\end{bmatrix}$$. 
	
	\item Alice and Bob locally compute $\sharetwo{x^{(\alpha)}} \gets \sharetwo{p^{(\alpha)}} \oplus \sharetwo{c^{(\alpha-1)}}$.
	
\end{enumerate}

\begin{figure*}[h]
	\centering
	\includegraphics[scale=0.45]{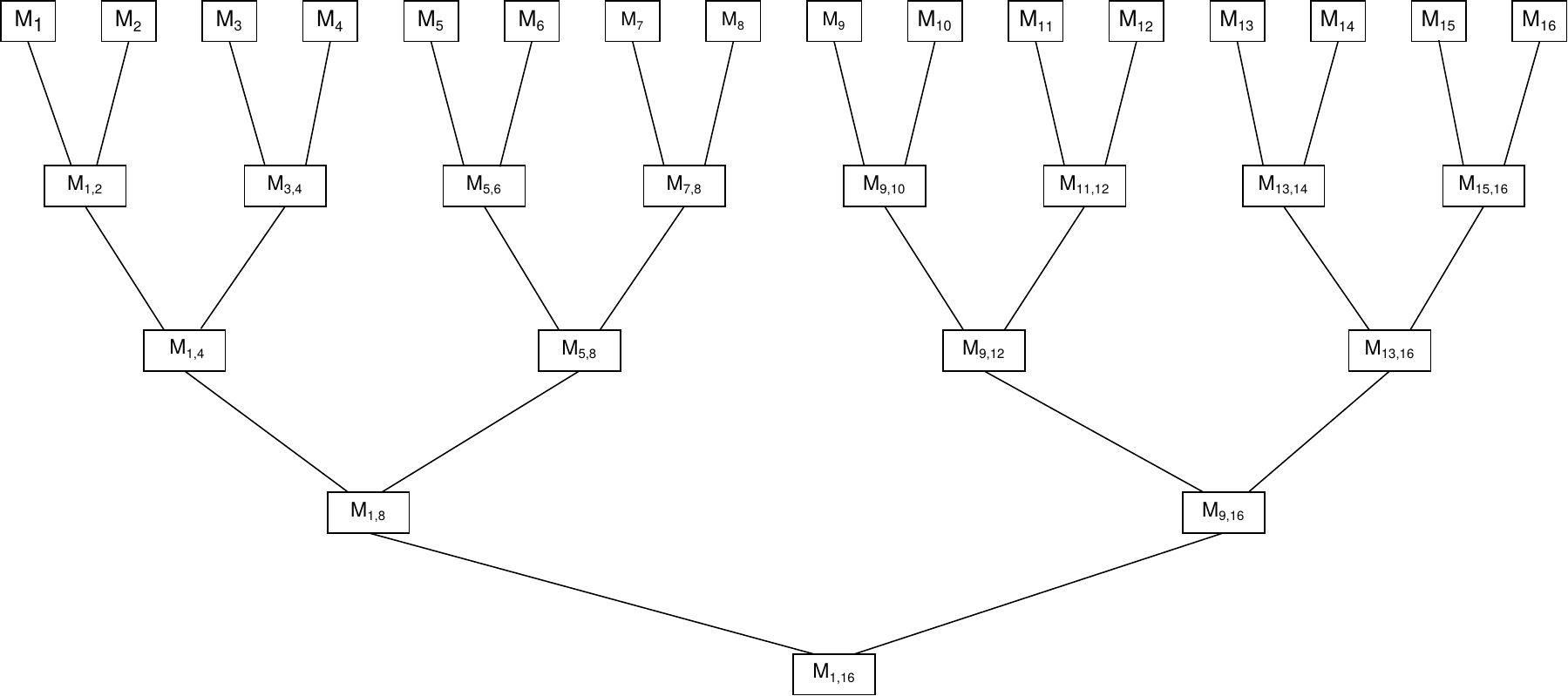}
	\caption{A circuit to compute the $(\alpha-1)$-th matrix composition in $\lceil \log(\alpha-1) \rceil$ layers. The notations $M_{i.j}$ indicates the composition of all matrices from $M_i$ to $M_j$, inclusive.}
	\label{fig:mat_composition_circuit_17}
\end{figure*}

The protocol $\pi_{\mathsf{{BTX}}}$ requires $\lceil{\log{\alpha-1)}}\rceil$ rounds and
$2(\alpha-1)+4\log{(\alpha-1)}-4$ bits of communication. Figure~\ref{fig:mat_composition_circuit_17} shows an example circuit to compute the matrix composition phase for $\alpha=17$. For batched inputs $\{ x_1, \dots, x_k \}$, the number of communication rounds remains the same and the total data transfer is scaled by $k$.

\noindent\textbf{Secure Comparison:} To perform a secure comparison of secret shared integers, we use the protocol $\pi_{\mathsf{GEQ}}$ of Adams \textit{et al.}~\cite{unpublishedDT}. As input, Alice and Bob hold secret shares in $\mathbb{Z}_q$ of integers $x$ and $y$ such that $|x-y|< 2^{\lambda-1}$ (as integers). Particularly, Alice and Bob can use this protocol with integers $x$ and $y$ in the range $[-2^{\lambda-2}, 2^{\lambda-2}-1]$ (a negative value $u$ is represented as $2^\lambda - |u|$). The protocol returns a secret share in $\mathbb{Z}_2$ of 1 if $x \geq y$ and of 0 otherwise. The protocol $\pi_{\mathsf{GEQ}}$ works as follows:

\begin{enumerate}
	
	\item Alice and Bob locally compute the difference of $x$ and $y$ as $\shareq{\mathsf{diff}} \leftarrow \shareq{x} - \shareq{y}$. Note that if $y > x$, then $\mathsf{diff}$ is negative. 
	
	\item Alice and Bob extract a $\mathbb{Z}_2$-sharing $\sharetwo{MSB}$ of the most-significant bit (MSB) of $\mathsf{diff}$ using the protocol $\pi_{\mathsf{{BTX}}}$. 
	
	\item Given that the most-significant bit of a secret shared value in $\mathbb{Z}_q$ is 1 if and only if it is negative, the negation of the most-significant bit, $\sharetwo{z} \leftarrow 1 + \sharetwo{MSB}$, is 1 if and only if $x \geq y$. 
	
\end{enumerate}

The protocol $\pi_{\mathsf{GEQ}}$ has the same round and communication complexity as the protocol $\pi_{\mathsf{{BTX}}}$ for extracting the most-significant bit, i.e., $\alpha=\lambda$. They differ solely on the computations done locally. 


	

\subsection{Naive Bayes Classifiers}\label{naive_bayes}

Naive Bayes is a statistical classifier based on Bayes' Theorem  with an assumption of independence among features/predictors. It assumes that the presence (or absence) of a particular feature in a class is unrelated to the presence (or absence) of any other feature. The  Bayes' theorem is used as follows: 
\begin{equation*}
P (c \vert x) = \frac{P(x \vert c) P(c)}{P(x)}, 
\end{equation*}
where: (1) $c$ is the class/category; (2) $x$ is the feature vector of test example; (3) $P( c \vert x)$ is the posterior probability, i.e., given test example $x$, what is its probability of belonging to class $c$; (4) $P(x \vert c)$ is known as the likelihood, i.e., given a class $c$, what is the probability of example $x$ belonging to class $c$; (5) $P(c)$ is the class prior probability; (6) $P(x)$ is the predictor prior probability.

The predictor prior probability $P(x)$ is the normalizing constant so that the $P(c\vert x)$ does actually fall in the range [0, 1]. In our solution we will be comparing the probabilities of different classes to determine the most likely class of an example. The probabilities are not important per se, only their comparative values are relevant. As the denominator $P(x)$ remains the same, it will be omitted and we will use
\[P(c \vert x) = P(x \vert c) P(c).\] 

As per assumption the features $x=(x_1,\ldots,x_d)$ are independent, we get
\[P(c \vert x)  = P(c) \prod^{d}_{k=1}P(x_k \vert c).\]

Note that when executing Naive Bayes, since the probabilities are often very small numbers, multiplying them will result in even smaller numbers, which often results in underflows that can cause the model to fail. To solve that problem and also simplify operations (and consequently improve performance), we will ``convert'' all multiplication operations into additions by using logarithms. Applying the logarithm we get
\begin{eqnarray*}
\log(P(c \vert x)) &=& \log\left(P(c)\prod^{d}_{k=1}P(x_k \vert c)\right)\\ 
&=& \log(P(c)) + \sum_{k=1}^{d} \log(P(x_k\vert c)). 
\end{eqnarray*}

To perform the classification, we then compute the argmax
\begin{equation*}\label{naive_bayes_classifier}
\hat{c} = \mathsf{argmax}_c\left[
\log(P(c)) + \sum_{k=1}^{d} \log(P(x_k \vert c))\right], 
\end{equation*}
where $\mathsf{argmax}_c$ returns the class $c$ that has the highest value for the test example $x$.

Naive Bayes classifiers differ mainly on the assumptions they make regarding the distribution of P($x \vert c$). In Gaussian Naive Bayes, the assumption is that the continuous values associated with each class are distributed according to a Gaussian distribution. For discrete features, as we have in text classification, we can use the Bernoulli Naive Bayes or the multinomial Naive Bayes. In the Bernoulli Naive Bayes, the features are Boolean variables, where 1 means that the word occurred in the text and 0 if the word did not occur. And, in the multinomial Naive Bayes, the features are the frequency of the words present in the document.
In this work, we use the multinomial Naive Bayes and the frequencies of the words are determined during the training phase.

\section{Privacy-Preserving Naive Bayes Classification}\label{PPNBC}
	
For the construction of our Privacy-Preserving Naive Bayes Classification (PPNBC) protocol $\pi_{\mathsf{PPNBC}}$, Alice constructs her set $ A= \{a_1, \ldots, a_m \}$ of unigrams occurring in her message and Bob constructs his set $B = \{b_1, \ldots, b_n\}$ of unigrams that occur in his ML model. Bob also has $\log(P(c_j))$, that is the logarithm of the probability for each class $c_j$ and a set of logarithms of  probabilities $\{\log(P(b_1 \vert c_j)), \ldots, \log(P(b_n \vert c_j)), \log(1-P(b_1 \vert c_j)), \ldots, \log(1-P(b_n \vert c_j))\}$, that is the logarithm of the probability of a word $b_i$ occurring or not in a class $c_j$. All $a_k$ and $b_i$ are represented as bit strings of length $\ell$. In our current implementation, we focus on binary classification. It is straightforward to generalize our protocols to the case of classification into more than two classes by using a secure argmax protocol. Our protocol $\pi_{\mathsf{PPNBC}}$ follows the description of the Naive Bayes presented in Section~\ref{naive_bayes} (using logarithms and not using the normalizing constant), and works as follows:
	
\begin{enumerate}
	
	\item Alice and Bob execute the secure feature extraction protocol $\pfe$ with inputs $(a_1, \ldots, a_m)$ and $(b_1, \ldots, b_n)$, respectively. The output consists of secret shared values $\sharetwo{y_1}, \ldots, \sharetwo{y_n}$ in $\mathbb{Z}_2$, where $y_i=1$ if the word $b_i \in A$ and 0 otherwise;
	
	\item They use protocol $\pconv$ to convert $\sharetwo{y_1}, \ldots, \sharetwo{y_n}$ to $\shareq{y_1}, \ldots, \shareq{y_n}$, containing secret sharings of the same values in $\mathbb{Z}_q$;
	
	\item For each class $c_j$:
	
	\begin{enumerate}
			
		\item Using the method described in Section \ref{sec:secsharing}, Bob creates secret shares of his inputs $\log(P(c_j)), \log(P(b_1 \vert c_j)), \ldots, \allowbreak \log(P(b_n \vert c_j)), \allowbreak \log(1-P(b_1 \vert c_j)), \ldots, \log(1-P(b_n \vert c_j))$, which contain the logarithm of the class probability and the set of logarithms of the conditional probabilities;
	
		\item For $i=1, \ldots, n$, Alice and Bob use the protocol $\pi_{\mathsf{DMM}}$ to compute 
		$\shareq{w_i} \gets \shareq{y_i}\shareq{\log(P(b_i \vert c_j))}+
		(1-\shareq{y_i})\shareq{\log(1-P(b_i \vert c_j))}
		$;

		\item Alice and Bob locally compute $\shareq{u_j} \gets \shareq{\log(P(c_j))} + \sum_{i=1}^{n} \shareq{w_i}$.
	
	\end{enumerate} 
	
	\item Alice and Bob use the protocol $\pi_{\mathsf{GEQ}}$ to compare the results of Step 3(c) for the two classes, getting as output a secret sharing of the output class $\sharetwo{c}$ (the secret sharing $\sharetwo{c}$ can afterwards be opened towards the party/parties that should receive the result of the classification).
	
\end{enumerate}

\section{Security}

\change{The concept of simulation is central in modern cryptography. It is used, for instance, to define zero-knowledge proofs, to appropriately analyze secure multi-party computation protocols, and is also behind the concept of semantic security for encryption schemes.
}

\change{At a high-level, in the simulation paradigm for security definitions and proofs, a comparison is made between a ``real world'' where the actual primitive being analyzed exists and the adversary tries to attack it, and an ``ideal world'' where there is an idealized primitive (also known as ideal functionality) that performs the desired functionality and is secure by definition. If one can prove that for any possible action taken by any adversary in the real world there is a corresponding action that an ideal-world adversary (also known as the simulator)  interacting with the ideal world can take such that the real and ideal worlds become indistinguishable, then the actual primitive securely realizes what is specified by the ideal functionality. In the case of semantic security for encryption schemes, for instance, one compares what can be learned by an adversary that receives a ciphertext in the real world with what can be learned by a simulator who receives nothing in the ideal world. An encryption scheme is semantically secure if the adversary cannot learn more information than the simulator. For a tutorial of the simulation proof technique, we refer to the work of Lindell \cite{eprint:lindell16}. Simulation-based security proofs generally offer stronger security guarantees than security proofs based on list of properties (and multi-party cryptographic protocols proved secure according to simulation-based definitions are sequentially composable \cite{JC:Canetti00}). 
}

\change{The security model considered in this work is the Universal Composability (UC) framework of Canetti \cite{Canetti01Uni}. 
The UC framework is based on the simulation paradigm, and thus in the UC framework the security is analyzed by comparing a real world with an ideal world. However, instead of considering the analyzed primitive isolated from the outside world (like other simulation-based definitions), the outside world is taken into account. By taking this additional factor into account, cryptographic protocols that are proven to be UC-secure offer much stronger security guarantees: any protocol that is proven UC-secure can be arbitrarily composed with other copies of itself and of other protocols (even with arbitrarily concurrent executions) while preserving security. That is an extremely useful property that allows the modular design of cryptographic protocols: consider the case in which one has previously proven that a protocol $\rho$ UC-securely realizes an ideal functionality $\mathcal{G}$. Now, consider that one has designed a new protocol $\pi$ that uses $\rho$ as a sub-protocol and wants to prove that $\pi$ securely realizes the ideal functionality $\F$. Due to the UC theorem \cite{Canetti01Uni}, in the security proof that protocol $\pi$ realizes functionality $\F$, one can consider $\pi$ using instances of the ideal functionality $\mathcal{G}$ (that is UC-realized by $\rho$) instead of instances of $\rho$; and this makes the modular design and security analysis of cryptographic protocols much easier. Moreover, UC-security is a necessity for cryptographic protocols running in complex environments such as the Internet. For these reasons, the UC framework is the gold standard for formally defining and analyzing the security of cryptographic protocols (it has been used by thousands of scientific works); and protocols that are UC-secure provide much stronger security guarantees than protocols proven secure according to other notions.
}

\change{Here only a short overview of the UC framework for the specific case of protocols with two participants is presented. We refer interested readers to the full version of original work of Canetti \cite{eprint:Canetti00} and the book of Cramer et al. \cite{CDN2015} for more details. }

\change{As mentioned before, the UC framework considers a real and an ideal worlds. In the real world Alice and Bob interact between themselves and with an adversary $\adv$ and an environment $\env$. The environment $\env$ captures all external activities to the protocol instance under consideration (i.e., it captures the outside world, everything other than the single protocol instance whose security is being analyzed), and is responsible for giving the inputs and getting the outputs from Alice and Bob.}
The adversary $\adv$ can corrupt either Alice or Bob, in which case he gains the control over that participant. The network scheduling is assumed to be adversarial and thus $\adv$ is responsible for delivering the messages between Alice and Bob. In the ideal world, there is an ideal functionality $\F$ that captures the perfect specification of the desired outcome of the computation. $\F$ receives the inputs directly from Alice and Bob, performs the computations locally following the primitive specification and delivers the outputs directly to Alice and Bob. A protocol $\pi$ executed between Alice and Bob in the real world UC-realizes the ideal functionality $\F$ if for every adversary $\adv$ there exists a simulator $\s$ such that no environment $\env$ can distinguish between: (1) an execution of the protocol $\pi$ in the real world with the participants Alice and Bob, and the adversary $\adv$; (2) and an ideal execution with dummy parties (that only forward inputs/outputs), $\F$ and $\s$.

\textbf{Simplifications:} The messages of ideal functionalities are formally public delayed outputs, meaning that $\s$ is first asked whether they should be delivered or not (this is due to the modeling that the adversary controls the network scheduling). This detail as well as the session identifications are omitted from the description of functionalities presented here for the sake of readability. 

\change{\textbf{Simulation Strategy:} As standard in UC security proofs, in our security proofs the simulator $\s$ interacting in the ideal world will internally run a copy of the adversary $\adv$ and internally simulate an execution of the protocol $\pi$ for $\adv$ (using only the information that $\s$ can get in the ideal world). The simulator $\s$ forwards the messages of $\adv$ and $\env$ that are intended to each other, thus allowing them to freely communicate (note that in the real world $\adv$ and $\env$ can freely communicate, so $\s$ should also allow this communication). One of the goals of the simulator is to make the internally simulated execution of the protocol $\pi$ and the real execution of the protocol $\pi$ in the real world indistinguishable from the point of view of $\adv$ and $\env$. Moreover, in the simulated execution of $\pi$ in the ideal world, $\s$ needs to extract the inputs of the corrupted parties in order to forward them to $\F$, and also make sure that the outputs in the simulated execution of $\pi$ and in the ideal functionality $\F$ match (note that the environment $\env$ can see the inputs/outputs of the uncorrupted parties directly from the dummy parties that simply forward inputs/outputs between $\F$ and $\env$). As long as the internally simulated and real executions of $\pi$ are indistinguishable and the inputs/outputs of both worlds match, the environment will not be able to distinguish the real and ideal worlds.}

\change{In the case of our (sub-)protocols all the computations are performed using secret sharings and all the protocol messages look uniformly random from the point of view of the receiver, with the single exception of the openings of the secret sharings. Nevertheless, the messages that open a secret sharing can be straightforwardly simulated using the outputs of the respective functionalities. In the ideal world, if $\pi$ uses $\rho$ as a sub-protocol and its has been previously shown that $\rho$ UC-realizes functionality $\mathcal{G}$, then using the UC theorem \cite{Canetti01Uni} it is possible to substitute the instances of $\rho$ used in $\pi$ by instances of $\mathcal{G}$.
And, in the ideal world, the simulator $\s$ has the leverage of being the one responsible for simulating all the ideal functionalities other than the one whose security is being analyzed. Using this leverage, the simulator $\s$ will be easily able to perform a perfect simulation in the case of our protocols.
}

\change{As shown in \cite{Dowsley16,IEEETDSC:CDHK+19}, 
the protocol $\pi_{\mathsf{DMM}}$ for secure matrix multiplication  UC-realizes the distributed matrix multiplication functionality $\fmmul$ in the trusted initializer model. The correctness follows trivially as $
Z = XY = (U+D)(V+E) = UV + UE + DV + DE = W + UE + DV + DE
$
and therefore $\shareq{Z} \gets \shareq{W} + E \shareq{U} + D \shareq{V} +DE$ obtains a secret sharing corresponding to $Z=XY$. The fact that the resulting shares are uniformly random with the constraint that $Z=XY$ follows trivially from the fact that the 
pre-distributed multiplication triple has this property. The simulator $\s$ runs internally a copy of the adversary $\adv$ and perfectly reproduces an execution of the real world protocol for $\adv$: $\s$ simulates an execution of $\pi_{\mathsf{DMM}}$ with 
dummy inputs for the uncorrupted parties (note that from $\adv$'s point of view the generated messages will be indistinguishable from the messages in the real protocol execution as the shares of $U$ and $V$ are uniformly random and unknown to $\adv$), and uses the leverage of being responsible for simulating the trusted initializer functionality $\fti{}$ for $\adv$ in order to extract the shares of $X$ and $Y$ whenever a corrupted party announces its shares of $D$ and $E$ in the simulated protocol execution. Having extracted the inputs of the corrupted party, $\s$ can forward them to the distributed matrix multiplication functionality $\fmmul{}$. Given the knowledge of $\s$ about $\shareq{U}, \shareq{V}, \shareq{W}, D$ and $E$, by the end of the simulated execution, it knows, for each corrupted party, the value that its share of the output should be, and therefore $\s$ can fix these values in $\fmmul{}$ so that the sum of the uncorrupted parties' shares is compatible with the simulated execution of $\pi_{\mathsf{DMM}}$. Given these facts, no environment $\env$ can distinguish the real and ideal worlds.
}

\begin{functionality}{Functionality $\fmmul{}$}
$\fmmul$ is parametrized by the size $q$ of the ring $\mathbb{Z}_q$ and the dimensions $(i, j)$ and $(j, k)$ of the matrices.\\
\\
\textbf{Input:} Upon receiving a message from Alice/Bob with its shares of $\shareq{X}$ and $\shareq{Y}$, verify if the share of $X$ is in $\Zqm{i}{j}$ and the share of $Y$ is in $\Zqm{j}{k}$.
If it is not, abort. Otherwise, record the shares, ignore any subsequent message from that party and
inform the other party about the receipt.\\

\textbf{Output:} Upon receipt of the shares from both parties, reconstruct $X$ and $Y$ from 
the shares, compute $Z=X Y$ and create a secret sharing $\shareq{Z}$ to distribute to Alice and Bob: a corrupt party fixes its share of the output to any chosen matrix and the shares of the uncorrupted parties are then created by picking uniformly random values subject to the correctness constraint.
\end{functionality}

\change{As proved by Reich \textit{et al.}~\cite{NeurIPS2019}, the protocol $\peq$ UC-realizes the functionality $\feq$. The correctness of $\peq$ follows trivially from the fact that in the case that $x=y$, then all $r_i$'s will be equal to 1 and therefore $z=\prod_i r_i$ will also be 1; and if $x\neq y$, then for at least one value $i$, we have that $r_i=0$, and therefore $z=0$. As showed by Reich \textit{et al.}~\cite{NeurIPS2019}, in the ideal world, $\s$ can run an internal copy of $\adv$ and a simulated execution of $\peq$ with dummy inputs (from $\adv$'s point of view the messages will be indistinguishable from the ones in the real protocol execution). Since $\pi_{\mathsf{DMM}}$ is substituted by $\fmmul$ using the UC composition theorem, and $\s$ is the one responsible for simulating $\fmmul$ in the ideal world, $\s$ can leverage this fact in order to extract the share that any corrupted party have of the value $x_i+y_i$. Let the extracted value be denoted by $v_{i,C}$. $\s$ then picks uniformly random $x_{i,C},y_{i,C} \in \{0,1\}$ such that $x_{i,C}+y_{i,C}=v_{i,C} \mod 2$ and submits them to $\feq$ as being the corrupted party's
shares of $x_i$ and $y_i$ (note that $\feq$'s output only depends on the values of $x_i+y_i \mod 2$). $\s$ is also trivially able to fix in $\feq$ the output share of the corrupted party so that it matches the one in the internally simulated instance of $\peq$. This is a perfect simulation strategy and no environment $\env$ can distinguish the ideal and real worlds. Therefore $\peq$ UC-realizes $\feq$.
}

\begin{functionality}{\textbf{Functionality} $\feq{}$}
{
$\feq$ is parametrized by the bit-length $\ell$ of the values being compared.\\

\textbf{Input:} Upon receiving a message from Alice/Bob with her/his shares of $\sharetwo{x_i}$ and $\sharetwo{y_i}$ for all $i \in \{1, \ldots, \ell\}$, 
record the shares, ignore any subsequent messages from that party and inform the other party about the receipt.\\
 
\textbf{Output:} Upon receipt of the inputs from both parties, reconstruct $x$ and $y$ from the bitwise shares. If $x = y$, then create and distribute to Alice and Bob the secret sharing $\sharetwo{1}$; otherwise the secret sharing $\sharetwo{0}$. Before the deliver of the output shares, a corrupt party fix its share of the output to any constant value. In both cases the shares of the uncorrupted parties are then created by picking uniformly random values subject to the correctness constraint.
}
\end{functionality}

\change{From the fact that $\peq$ UC-realizes $\feq$, it follows straightforwardly that $\pfe$ UC-realizes the functionality $\ffe$. The correctness is trivial to verify and $\feq$ does not reveal any information at all about the secret shared values. In the ideal world, 
$\s$ executes an internal copy of $\adv$ and simulates an execution of the protocol $\pfe$ for $\adv$. Note that in this internal simulation $\s$ can use the leverage of being responsible for simulating $\fti{}$ in order to extract all inputs of the corrupted party, which can then be forwarded to $\ffe$. $\s$ can also fix in $\ffe$ the output shares of the corrupted party to match the ones in the internally simulation execution. No environment $\env$ is able to distinguish the real and ideal worlds, and thus $\pfe$ UC-realizes the functionality $\ffe$.
}

\begin{functionality}{\textbf{Functionality} $\ffe$}
{
$\ffe$ is parametrized by the sizes $m$ of Alice's set and $n$ of Bob's set, and the bit-length $\ell$ of the elements.\\

\textbf{Input:} Upon receiving a message from Alice with her set $A= \{a_1,a_2,\ldots,a_m\}$ or from Bob with his set $B= \{b_1,b_2,\ldots,b_n\}$, record the set, ignore any subsequent messages from that party and inform the other party about the receipt.\\

\textbf{Output:} Upon receipt of the inputs from both parties, define the binary feature vector $x$ of length $n$ by setting each element $x_i$ to $1$ if $b_i \in A$, and to $0$ otherwise. Then create and distribute to Alice and Bob the secret sharings $\sharetwo{x_i}$. Before the deliver of the output shares, a corrupt party fix its share of the output to any constant value. In both cases the shares of the uncorrupted parties are then created by picking uniformly random values subject to the correctness constraint.
}
\end{functionality}

\change{As proved by Reich \textit{et al.}~\cite{NeurIPS2019}, the protocol $\pconv$ UC-realizes the functionality $\fconv$. The correctness of $\pconv$ is trivial to verify: as $x=x_a+x_B \mod 2$, then $z=x_A+x_B-2x_Ax_B$ is such that $z=x$ for all possible values $x_A, x_B \in \{0,1\}$. In the internal simulation of $\pconv$ in the ideal world, $\s$ can use the fact that it is the one simulating $\fmmul$ in order to extract the share of any corrupted party and fix the input to/output from $\fconv$ appropriately, so that no environment $\env$ can distinguish the real and ideal worlds. Hence $\pconv$ UC-realizes $\fconv$.
}

\begin{functionality}{\textbf{Functionality} $\fconv$}
{
$\fconv$ is parametrized by the size of the field $q$.\\

\textbf{Input:} Upon receiving a message from Alice/Bob with her/his share of $\sharetwo{x}$, record the share, ignore any subsequent messages from that party and inform the other party about the receipt.\\

\textbf{Output:} Upon receipt of the inputs from both parties, reconstruct $x$, then create and distribute to Alice and Bob the secret sharing $\shareq{x}$. Before the deliver of the output shares, a corrupt party fix its share of the output to any constant value. In both cases the shares of the uncorrupted parties are then created by picking uniformly random values subject to the correctness constraint.
}
\end{functionality}

\change{The bit extraction protocol $\pi_{\mathsf{{BTX}}}$ of \cite{unpublishedDT} is a straightforward simplification of the bit decomposition protocol \pdecompopt from \cite{Cock20} and UC-realizes the bit extraction functionality $\fbtx$. Note that the simulator $\s$ can straightforwardly extract the bit-string of a corrupted party in an internal simulation of $\pi_{\mathsf{{BTX}}}$ with the adversary $\mathcal{A}$ by using the fact that it is responsible for simulating $\fmmul$ that is used to compute the generate signal. Thus $\s$ can forward the necessary inputs $\fbtx$. It can also easily fix the output share of the corrupted party in $\fbtx$ so that it matches the one in the internal simulation of protocol $\pi_{\mathsf{{BTX}}}$.
Therefore $\s$ has a perfect simulation strategy and $\env$ cannot distinguish the ideal and real worlds.}

\begin{functionality}{Functionality $\fbtx$}
$\fbtx$ is parametrized by $\alpha$. It receives bit-strings $x_A=x_A^{(\alpha)}\cdots x_A^{(1)}$ and 
$x_B=x_B^{(\alpha)}\cdots x_B^{(1)}$ from Alice and Bob, respectively, and returns a secret sharing of the $\alpha$-th bit of $x=x_A+x_B$.\\
\\
\textbf{Input:} Upon receiving a message from Alice with her bit-string $x_A$ or from Bob with his bit-string $x_B$, record it, ignore any subsequent messages from that party and inform the other party about the receipt.\\
\\
\textbf{Output:} Upon receipt of the inputs from both parties, 
compute $x=x_A+x_B$, extract the $\alpha$-th bit $x_\alpha$ of $x$ and distribute a new secret sharing $\sharetwo{x_\alpha}$ of the bit $x_\alpha$. Before the output deliver, the corrupt party fix its shares of the output to any desired value. The shares of the uncorrupted parties are then created by picking uniformly random values subject to the correctness constraints.
\end{functionality}

\change{Proceeding to the analysis of protocol $\pi_{\mathsf{GEQ}}$, its correctness follows trivially. In the ideal world, the simulator $\s$ executes an internal copy of $\mathcal{A}$ interacting with an instance of protocol $\pi_{\mathsf{GEQ}}$ in which the uncorrupted parties use dummy inputs. Note that all the messages that $\mathcal{A}$ receives look uniformly random to him. Since $\pi_{\mathsf{{BTX}}}$ is substituted by $\fbtx$ using the UC composition theorem, and $\s$ is responsible for simulating $\fbtx$ in the ideal world, $\s$ can leverage this fact in order to extract the share that any corrupted party have of the secret shared value $\mathsf{diff}=x-y$. Let the extracted value of the corrupted party be denoted by $\mathsf{diff}_C$. The simulator then picks uniformly random values $x_C, y_C$ in $\mathbb{Z}_q$ such that $\mathsf{diff}_C=x_C-y_C$ and submit these values to
$\fgeq$ as being the shares of the corrupted party for secret shared values $x$ and $y$ (note that the result of $\fgeq$ only depends on the value of $x-y \mod q$). 
$\s$ is also trivially able to fix the output share of the corrupted party in $\fgeq$ so that it matches the one in the internally simulated instance of $\pi_{\mathsf{GEQ}}$. This is a perfect simulation strategy and no environment $\env$ can distinguish the ideal and real worlds. Therefore $\pi_{\mathsf{GEQ}}$ UC-realizes $\fgeq$.
}

\begin{functionality}{Functionality $\fgeq$}
$\fgeq$ runs with Alice and Bob and is parametrized by the bit length $\lambda$ of the ring $\mathbb{Z}_q$ (i.e., $q=2^\lambda$). It receives as input the secret shared values $x$ and $y$, which are guaranteed to be such that $|x-y|< 2^{\lambda-1}$ (as integers).
\\
\\
\textbf{Input:} Upon receiving a message from Alice or Bob with its share of $\shareq{x}$ and $\shareq{y}$, record the shares, ignore any subsequent messages from that party and inform the other party about the receipt.\\
\\
\textbf{Output:} Upon receipt of the inputs from both parties, reconstruct the values $x$ and $y$, 
and compute $\mathsf{diff}=x-y$. If $\mathsf{diff}  $ represents a negative number, distribute a new secret sharing $\sharetwo{0}$; otherwise a new secret sharing $\sharetwo{1}$. Before the output deliver, the corrupt party fix its shares of the output to any desired value. The shares of the uncorrupted parties are then created by picking uniformly random values subject to the correctness constraints.
\end{functionality}

\change{When we analyze the security of the final protocol $\pi_{\mathsf{PPNBC}}$ for privacy-preserving Naive Bayes classification, we can use the UC theorem \cite{Canetti01Uni} to substitute the instances of $\pfe$, $\pconv$, $\pi_{\mathsf{DMM}}$ and $\pi_{\mathsf{GEQ}}$ that are used as sub-protocols in 
$\pi_{\mathsf{PPNBC}}$ by instances of $\ffe$, $\fconv$, $\fmmul$ and $\fgeq$, respectively. Note that these ideal functionalities do not leak any information at all to the protocol participants: the functionalities only manipulate the secret sharings of the inputs in order to obtain secret sharings of the desired outputs, but no secret shared value is ever opened to any party. Thus from the point of view of any protocol participant, all values that it sees throughout the execution of $\pi_{\mathsf{PPNBC}}$ look uniformly random. In the ideal world, $\s$ internally runs a copy of $\adv$ and simulates an execution of the real world protocol $\pi_{\mathsf{PPNBC}}$ for $\adv$. Using the leverage of being responsible for simulating $\fti{}$, $\ffe$, $\fconv$, $\fmmul$ and $\fgeq$ in this internal simulation of the protocol $\pi_{\mathsf{PPNBC}}$ for the adversary $\mathcal{A}$, the simulator $\s$ is trivially able to extract all inputs of the corrupted party that it needs to forward to the functionality $\fppnbc$ in this ideal world. In other words, $\s$ can extract $(a_1, \ldots, a_m)$ in the case Alice is corrupted; and it can extract $(b_1, \ldots, b_n)$ and $(\log(P(c_j)), \log(P(b_1 \vert c_j)), \ldots, \log(P(b_n \vert c_j)), \allowbreak
\log(1-P(b_1 \vert c_j)), \ldots, \log(1-P(b_n \vert c_j)))$ for each class $c_j$ in the case that Bob is corrupted. Knowing all values of the internal simulation, $\s$ can also trivially fix the share of the output that corresponds to the corrupt party in order to match the one from the internal execution. Therefore, it follows straightforwardly that $\pi_{\mathsf{PPNBC}}$ UC-realizes functionality $\fppnbc$.
}

\begin{functionality}{Functionality $\fppnbc$}
\textbf{Input:} Upon receiving a message from Alice with her 
inputs $(a_1, \ldots, a_m)$ or from Bob with his inputs $(b_1, \ldots, b_n)$ and $(\log(P(c_j)), \log(P(b_1 \vert c_j)), \ldots, \log(P(b_n \vert c_j)), \allowbreak
\log(1-P(b_1 \vert c_j)), \ldots, \log(1-P(b_n \vert c_j)))$ for each class $c_j$, record the values, ignore any subsequent messages from that party and inform the other party about the receipt.\\
 \\
\textbf{Output:} Upon receipt of the inputs from both parties, locally perform the same  computational steps as $\pi_{\mathsf{PPNBC}}$ using the secret sharings. Let $\sharetwo{c}$ be the result. Before the deliver of the output shares, a corrupt party can fix the shares that it will get, in which case the other shares are adjusted accordingly to still sum to $\sharetwo{c}$. The output shares are delivered to the parties.
\end{functionality}

\section{Experimental results}

\begin{figure*}[!h]
	\begin{center}
		\includegraphics[width=0.65\textwidth]{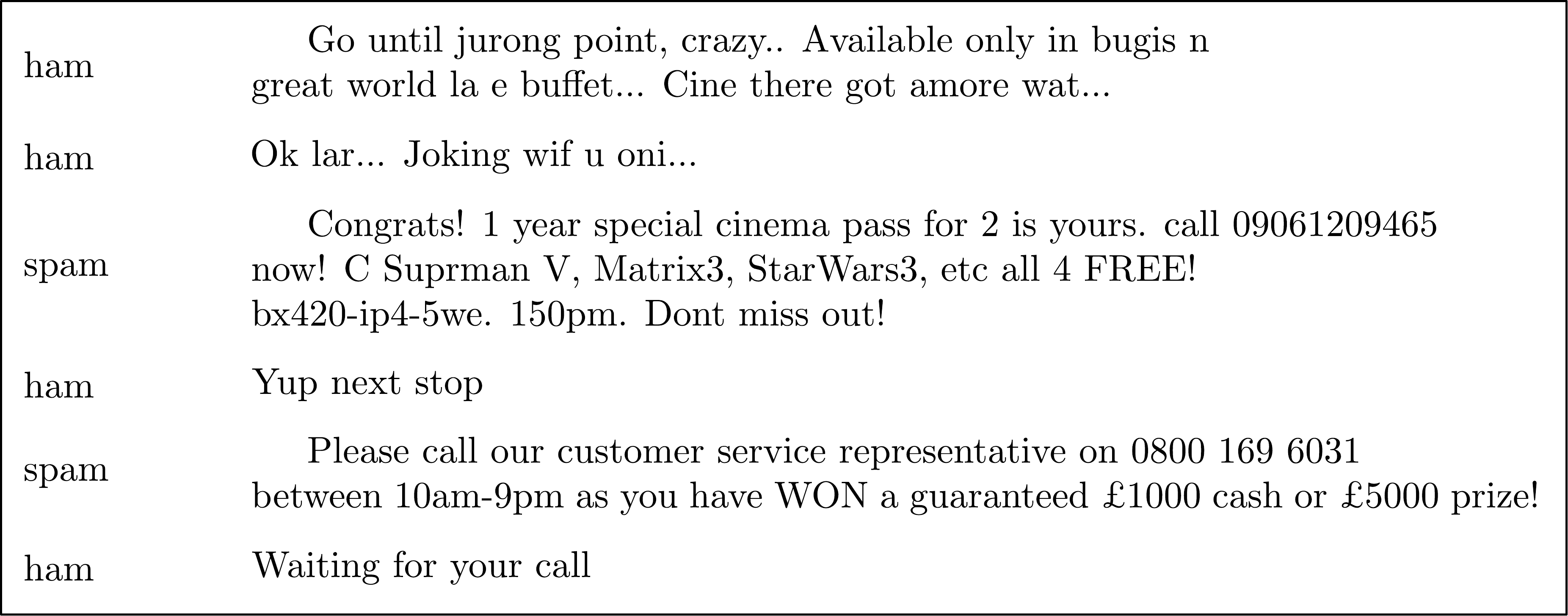}
		\caption{Some examples of tagged SMS messages from the SMS Spam Collection Data Set.}
		\label{dataset}       	
	\end{center}
\end{figure*}

To evaluate the proposed protocol in a use case for spam detection, we use the SMS Spam Collection Data Set from the UC Irvine Machine Learning Repository.\footnote{\url{https://archive.ics.uci.edu/ml/datasets/sms+spam+collection}} 
This database contains a set of tagged SMS messages that have been collected for SMS spam research. It contains a set of 5574 SMS messages in English tagged as legitimate (ham) or spam. The files contain one message per line, where each line is composed of two columns: the first contains the label (ham or spam) and the second contains the raw text (see examples in Figure~\ref{dataset}). The data set has 747 spam SMS messages and 4827 ham SMS messages, that is, 13.4\% of the SMSes are spams and 86.6\% are hams.
	
Table~\ref{token} shows the distribution of tokens/unigrams in the data set. As we can see, the data set has a total of 81175 tokens. When training a spam classifier, techniques can be used to reduce this set of tokens in order to improve the performance of the protocol in terms of accuracy, runtime or other metrics.
	
\begin{table}[!h]
	\begin{center}
	    \caption{Token statistics~\cite{Almeida11}.}\label{token}
		\begin{tabular}{|l|r|}
				\hline
			Tokens in Hams &  63632 \\ \hline
				
			Tokens in Spams & 17543 \\ \hline
				
			Total of Tokens & 81175 \\ \hline
			
			Average Tokens per Ham & 13.18 \\ \hline
				
			Average Tokens per Spam & 23.48 \\ \hline
			
			Average Tokens per Message & 14.56 \\  \hline
				
		\end{tabular}

	\end{center}
\end{table}

\subsection{Training phase}
	
In the classification phase, Bob already has the ML model. The model was generated using the following steps:
	
\begin{enumerate}
		
	\item Bob takes the SMS Spam Collection Data Set and parses each line into unigrams.
	The letters are converted to lower case and everything other than letters is deleted.
		
	\item To have higher accuracy and improve the runtime of the algorithm, we used the stemming and stop words techniques. Stemming is the process of reducing the inflection of words in their roots, such as mapping a group of words to the same stem even if the stem itself is not a valid word in the language. For example, likes, liked, likely and liking reduce to the stem like; retrieval, retrieved, retrieves reduce to the stem retrieve; trouble, troubled and troubles reduce to the stem troubl. Stop words concerns filtering out words that can be considered irrelevant to the classification task such as: the, a, an, in, to, for.
		
	\item The remaining unigrams are inserted in a Bag of Words (BoW). A BoW is created for 
	the ham category and another for the spam category. Each BoW contains the unigrams and their corresponding frequency counters.
		
	\item Based on the frequency counters, we remove the less frequent words in order to decrease the runtime of our privacy-preserving solution. We will address this parameter later when we detail the trade-off between accuracy and efficiency.
		
	\item Bob computes the logarithm of the class prior probability for each class $c$:
	\begin{equation}\label{prior_training}
	\log(P(c)) = \log\left(\dfrac{\vert \text{training examples} \in c \vert}{\vert \text{examples in the training set}\vert }\right).
	\end{equation}
	
	\item Bob computes the logarithm of the probability of each word by class. To compute the probability we have to find the average of each word for a given class. For the class $c$ and the word $i$, the average is given by:
	
	\begin{equation}
	\log(P(i \vert c)) = \log\left(\dfrac{\vert 
	\text{word ``i'' in class }  c \vert}{\vert
	\text{words in class } c \vert} \right).
	\end{equation}
	
	However, as some words can have 0 occurrences, we use Laplace Smoothing:
	
	\begin{equation}\label{likehood_training}
	\log(P(i \vert c)) = \log\left(\dfrac{\vert \text{word ``i'' in class }  c \vert +1}{\vert \text{words in class } c \vert + \vert V \vert}\right),
	\end{equation}
	
	where $\vert V \vert$ is the size of the vocabulary, i.e., all unique words in the training data set regardless of the class.
	
\end{enumerate}

In Equations~\ref{prior_training} and~\ref{likehood_training}, before computing the logarithm we need to scale the result of the division to convert it to integers. Before inputing this model into our privacy-preserving protocol, we need to convert any fixed precision real number into an integer. In order to do so, we follow section \ref{section:fixed_point_arithmetic}, we pick up a value of $a$ equal to 34. With the values of $\log(P(c))$ and $\log(P(i \vert c))$ computed, the model is generated. Note that only Bob is involved in training the model.

Table~\ref{token_training} shows the distribution of tokens/unigrams in the data set after performing the training phase. Compared to the Table~\ref{token}, there was a reduction of over 30 thousand tokens. Besides, we can see that we have less than 6000 unique tokens.
		
\begin{table}[!h]
\caption{Token statistics after the training phase.}\label{token_training}
	\begin{center}
	
		\begin{tabular}{|l|r|}
		\hline
			Tokens in Hams &  38469  \\ \hline
			Tokens in Spams & 10981 \\ \hline
			Total of Tokens & 49450 \\ \hline
			Average Tokens per Ham & 7.97 \\ \hline
			Average Tokens per Spam & 14.7 \\ \hline
			Average Tokens per Message & 8.87 \\ \hline
			Unique Tokens in Hams & 5950 \\ \hline
			Unique Tokens in Spams & 1883 \\ \hline
			Unique Tokens in the Data Set & 5950 \\ \hline			
		\end{tabular}
	\end{center}
\end{table}

\subsection{Cryptographic Engineering}

Our solution to secure Naive Bayes classification is implemented in Rust using an up-to-date version of the RustLynx framework (available at \url{https://bitbucket.org/uwtppml/rustlynx/src/master/}), which was used in \cite{Cock20} to achieve a state-of-the-art implementation of secure logistic regression training. The supported primitives are, to the best of our knowledge, the fastest available for two-party computation in the honest-but-curious setting when performed in a local area network. 

\paragraph{Parallel Data Transfer}

Instead of atomic sending and receiving queues as might be utilized in a general-purpose multi-threaded network application, we associate each thread (for a fixed threadpool size) with a port in a given port range such that the $i$-th thread executing in the $P_0$ process will only exchange data with the $i$-th thread of the $P_1$ process. We base this choice on the observation that MPC operations are symmetric, so there is never an instance where the receiver does not know the length, intent, or timing of a message from the sender -- situations for which a more complex, and slower, messaging system would be necessary. An additional benefit of this structure is that the packets require no header to denote the sender, thread ID, or length of the body. Based on empirical testing on $\mathbb{Z}_{2^{64}}$ multiplication with an optimized threadpool size, this method yields a 6$\times$ improvement over an architecture with atomic queues.

\paragraph{Operations in $\mathbb{Z}_{2^\lambda}$}

RustLynx supports arithmetic over $\mathbb{Z}_{2^{64}}$. This particular bit length is chosen because (1) it is sufficiently large to represent realistic data in a fixed-point form and (2) it aligns with a primitive data type, meaning that modular arithmetic operations can be performed implicitly by permitting integer overflow.

\paragraph{Operations in $\mathbb{Z}_2$}

We represent $\mathbb{Z}_2$ shares in groupings of 128 as the individual bits of Rust's unsigned 128-bit integer primitive. Doing so allows for local operations on the entire group of secrets to share Arithmetic Logic Unit (ALU) cycles and to be loaded, copied, and iterated quickly. The downside of this design choice is that sending $m$ $\mathbb{Z}_2$ shares corresponds to $16 \cdot \lceil m / 128 \rceil$ bytes of data transfer, which, in the worst case, is 15 bytes larger than the most compact possible representation of $m$ bits (that is, using groups of 8). Based on empirical testing, the performance loss for MPC primitives is affected significantly more by wasting time on local operations than wasting a small amount of bandwidth. So, the largest available primitive data type was chosen to group $\mathbb{Z}_2$ shares.

\subsection{Evaluation}

We ran our experiments on Amazon Web Services (AWS) using two c5.9x-large EC2 instances with 36 vCPUs, 72.0 GiB of memory and 32 threads. Each of the parties ran on separate machines (connected over a Gigabit Ethernet LAN), which means that the results in Table~\ref{runtime} cover the communication time in addition to the computation time. All experiments were repeated one-hundred times and averaged to minimize the variance caused by large thread counts.

We evaluate PPNBC using 5-fold cross validation over the entire corpus of 5574 SMS. For each unigram in Alice's set $A = \{a_1, a_2, \cdots, a_m\} $ and each unigram in Bob's set $ B = \{b_1, b_2, \cdots, b_n \} $, we apply the hash function SHA-256 (and truncate the result) to transform each one into a bit-string of size $\ell = 14$.

We evaluated our solution for $m = \{8, 16, 50, 160\}$ and $n = \{369, 484, 688, 5200 \}$. Note that the $n$ value affects the accuracy and running time. The $n$ values were defined based on the frequency of tokens appearing in the training data set: 688 tokens appeared more than 9 times; 484 tokens appeared more than 14 times; and 369 tokens appeared more than 19 times. We noticed a significant degradation of the False Positive Rate (FPR) when further reducing $n$.  The values of $m$ were defined based on the size of our messages. Note that $m$ determines the number of tokens in the message and not the number of characters. Also, we should mention that some messages in our data set consisted of multiple SMSes concatenated. Our maximum value of $n$ (160 tokens) is twice the maximum message found in our data set.  We recall that a single SMS has a 160 7-bit characters maximum. The average lengths found for SMS classified as ham or spam in our data set are shown in Table~\ref{token_training}.

We evaluate the proposed protocol in a use case for SMS spam detection, however our PPNBC can be used in any other scenario in which the Naive Bayes classifier can be employed. It is important to note that designing a model to obtain the highest possible accuracy is not the focus of this paper. Instead, our goal is to demonstrate that a privacy-preserving Naive Bayes classifier based on MPC is feasible in practice. Despite this, as shown in Table~\ref{acuracy}, the protocol achieves good results when compared to the best result presented by Almeida \textit{et al.}~\cite{Almeida11}, where the data set is proposed. They reach an accuracy equal to 97.64\%, a false positive rate (FPR) equal to 16.9\% and a false negative rate (FNR) equal to 0.18\% using a SVM classifier. In our best scenario ($n = 5200$), we have an accuracy equal to 96.8\%, FPR equal to 17.94\% and FNR equal to 0.87\%. We remark that there is little variation in accuracy and FNR when using smaller values of $n$. \change{Classifiers based on boosted decision trees (AdaBoost) and logistic regression, previously used in \cite{NeurIPS2019}, achieved accuracies within 0.5\% of the best accuracy achieved by our protocol.}

\begin{table}[h]
	\centering
		\caption{Accuracy results using 5-fold cross-validation over the corpus of 5574 SMS. FPR is the false positive rate and FNR is the false negative rate.}
	\label{acuracy}
	\begin{tabular}{|c|c|c|c|}		
		\hline
		Dictionary size  & FNR  & FPR & Accuracy  \\ \hline
		n=369  & 0.79\% & 28.52\% & 95.5\%  \\ \hline
		n=484  & 0.89\% & 22.22\% & 96.2\%  \\ \hline
		n=688  & 0.87\% & 21.15\% & 96.4\% \\ \hline
		n=5200 & 0.87\% & 17.94\% & 96.8\% \\ \hline		
	\end{tabular}
\end{table}

Table~\ref{runtime} reports the runtime of our PPNBC for different sizes of $m$ and $n$, where  $n$ is the size of the dictionary, that is, the amount of unigrams belonging to Bob's trained model and $m$ is the number of unigrams present in Alice's SMS. The feature vector extraction (Extr) runtime considers the time required to execute the Protocols $\pfe$ and $\pconv$ in the steps 1 and 2 of Protocol $\pi_{\mathsf{PPNBC}}$. The runtime for classification (Class) considers the remaining steps of the Protocol $\pi_{\mathsf{PPNBC}}$. And, the total runtime is Extr + Class. We can see that the runtime for classification is independent of the size of $m$, and is based only on the size $n$, and even for $n = 5200$ features/unigrams it only takes $48$\,ms. The feature vector extraction (Extr) runtime depends on both $m$ and $n$. For $n = 5200$ and $m = 160$, the feature extraction takes less than $290$\,ms, while for $n = 369$ and $m = 8$ it just takes $11$\,ms. The total runtime takes a maximum of $334$\,ms to classify a SMS with $160$ unigrams using a dictionary with $5200$ unigrams, and just $21$\,ms to classify a SMS with 8 unigrams using a dictionary with $369$ unigrams.

\begin{table*}[h]
	\centering
	\caption{Total runtime in milliseconds (Total) needed to securely classify a SMS with our proposal. We divided it in the time needed for feature vector extraction (Extr) and the time for classification (Class). $n$ is the size of the dictionary, that is, the amount of unigrams belonging to Bob's trained model and $m$ is the amount of unigrams present in Alice's SMS.}
	\label{runtime}
	\begin{adjustbox}{max width=\textwidth}
		\begin{tabular}{|l|r|r|r|r|r|r|r|r|r|r|r|r|}		\hline
			& \multicolumn{12}{c|}{\textbf{SMS size}} \\ \hline	
			\multirow{2}{*}{\pbox{\textwidth}{\textbf{Dictionary} \\ \textbf{size}}} & \multicolumn{3}{c}{m=8} & \multicolumn{3}{|c}{m=16} & \multicolumn{3}{|c}{m=50} &  \multicolumn{3}{|c|}{m=160} \\ \cline{2-13}		
			& Extr & Class & Total & Extr & Class & Total & Extr & Class & Total & Extr & Class & Total \\ \hline
			
			n=369   & 11\,ms & 10\,ms & 21\,ms & 25\,ms & 10\,ms &  35\,ms & 102\,ms & 10\,ms & 112\,ms & 111\,ms & 10\,ms & 121\,ms \\ \hline		
			n=484   & 12\,ms & 10\,ms & 22\,ms & 26\,ms & 10\,ms & 36\,ms & 103\,ms & 10\,ms &113\,ms & 124\,ms & 10\,ms & 134\,ms \\ \hline
			n=688	& 20\,ms & 11\,ms & 33\,ms & 36\,ms & 11\,ms & 47\,ms & 106\,ms & 11\,ms & 117\,ms & 136\,ms & 11\,ms & 147\,ms \\ \hline
			n=5200 	& 77\,ms & 48\,ms & 125\,ms & 89\,ms & 48\,ms & 137\,ms & 140\,ms & 48\,ms & 188\,ms & 286\,ms & 48\,ms & 334\,ms \\ \hline				
			
		\end{tabular}
	\end{adjustbox}
		
\end{table*}

As we can notice from Table~\ref{runtime}, the feature extraction is the part that spends the most time because $m \times n$ secure equality tests of bit strings are required, which are based on secure multiplications. As discussed by Reich \textit{et al.}~\cite{NeurIPS2019}, the number of secure equality tests needed could possibly be reduced if Alice and Bob first map using the same hash function each of their bitstrings to $t$ buckets, $A_1, A_2, \cdots, A_t$ for Alice and $B_1, B_2, \cdots, B_t$ for Bob. Then, only the bitstrings belonging to $A_i$ would need to be compared with the bitstrings belonging to $B_i$.

To use buckets, each $a_m$ and $b_n$ element is hashed, and the result is divided into two parts, where the first $q$ bits indicate which bucket the element belongs to and the other $r$ bits are stored, thus $t = 2^{q}$. To hide how many elements are mapped to each bucket, as this can leak information about the distribution of the elements, the empty spots of each bucket must be filled up  with dummy elements. Thus, considering $s_1$ the size of each bucket $A_t$ and $s_2$ the size of each bucket $B_t$, the extraction feature protocol will need $t \times s_1 \times s_2$ equality tests, which can be substantially smaller than $n \times m$ needed previously. It is important to note that these dummy elements do not modify  the accuracy (or any other metrics) of the classification, because when generating Bob's model, for each dummy element, the probability of the element to occur for each class is defined as 0, that is, it does not impact $\sum_{k = 1}^{d} \log(P(x_k \vert c))$. In our case, since the values of $m$ and $n$ are not large, there is no significant difference between using buckets or not. Therefore, we use the original version (without buckets), as in this case there is no probability of information being leaked due to buckets' overflow.

We remark that, for the sake of evaluating our solution, we have selected values of $n$ (the dictionary size) that directly depend on the frequency of tokens. That is not necessary in general. 

\change{\noindent\textbf{Communication and Round Complexities.} We provide an analysis of the communication and round complexities of our protocol. Let $n$ be the dictionary length, $m$ the example length, $\ell$ the hash output length, and $q=2^\lambda$ the ring order (remember that our sub-protocols use the binary field or a larger ring, depending on the operations). Our protocol $\pi_{\mathsf{PPNBC}}$ makes 1 call to the protocol $\pfe$, $n$ calls to $\pconv$ (which can be done in parallel) and one call to protocol $\pi_{\mathsf{GEQ}}$. Protocol $\pfe{}$ requires $\lceil \log(\ell) \rceil +1$ rounds of communication and $4 \cdot m \cdot n \cdot (\ell -1) + m + n$ bits of communication. Protocol $\pconv$ requires 1 round of communication and $4 \lambda$ bits of communication. Protocol 
$\pi_{\mathsf{GEQ}}$ requires $\lceil{\log{\lambda-1)}}\rceil$ rounds and
$2(\lambda-1)+4\log{(\lambda-1)}-4$ bits of communication. 
So, in total, protocol $\pi_{\mathsf{PPNBC}}$ requires $ \lceil{\log{\ell}} \rceil  + \lceil{\log{\lambda-1)}}\rceil + 2$ rounds and 
$4 \cdot m \cdot n \cdot (\ell -1) + m + n + 4 \cdot \lambda \cdot n + 2(\lambda-1)+4 \log{(\lambda-1)}-4$ 
bits of communication.
}

\section{Related work}

Privacy-preserving versions of ML classifiers were first addressed by Lindell and Pinkas~\cite{LindellP00}. They used MPC to build a secure ID3 decision tree where the training set is distributed between two parties. Most of the literature on privacy-preserving ML focus on the training phase, and include secure training of ML algorithms such as Naive Bayes~\cite{WrightY04,VaidyaKC08,CoudertM11}, decision tree~\cite{LindellP00,BrickellS09,FC:HSCA14}, logistic regression~\cite{ChaudhuriM08,mohassel2017secureml,Cock20}, linear regression~\cite{AISec:CDNN15,mohassel2017secureml,IEEENSRE:ADMW+19}, neural networks~\cite{ShokriS15,mohassel2017secureml,PoPETS:WagGupCha19} and SVM~\cite{VaidyaYJ08}. Regarding privacy-preserving classification/inference/prediction, most works focused on secure neural network inference, e.g.,  \cite{BarniFLS011,Gilad16,mohassel2017secureml,Liu17,Riazi18,Juvekar18,PoPETS:WagGupCha19,CCS:ASKG19,USENIX:RSCLLK19,PoPETS:DalEscKel20,USENIX:MLSZP20,SP:KRCGRS20,CCS:RRKCGRS20}.
Far less works focus on privacy in the classification/prediction phase of other algorithms.
De Hoogh et al. \cite{FC:HSCA14} has a protocol for privacy-preserving scoring of decision trees. Bost \textit{et al.}~\cite{BostPTG15} proposed privacy-preserving classification protocols for hyperplane-based classifiers, Naive Bayes and decision trees where the description of the features (the dictionary in our implementation) was supposed to be public. David et al. \cite{david2015efficient} presented protocols for privacy-preserving classification with hyperplane-based classifiers and Naive Bayes, again, the classifier features were supposed to be publicly known. Our solution guarantees the privacy of the dictionary. Khedr \textit{et al.}~\cite{KhedrGV16} proposed a secure NB classifier based on Fully Homomorphic Encryption (FHE). De Cock et al. \cite{IEEETDSC:CDHK+19} presented private scoring protocols for decision trees and hyperplane-based classifiers. Fritchman et al. \cite{fritchman2018} presented a solution for private scoring of tree ensembles.

Regarding privacy-preserving text classification, Costantino et al. \cite{CostantinoMMSS17} presented a proposal based on homomorphic encryption that takes 19 minutes to classify a tweet, with 19 features and 76 minutes to classify an email with 17 features. In addition to the high runtime, Bob learns which of his lexicon’s words are present in Alice’s tweets.

\change{To the best of our knowledge, the work by Reich et al.~\cite{NeurIPS2019}  was the first work to present solutions for privacy-preserving feature extraction and classification of unstructured texts based on secure multiparty computation. We present their results for dictionary sizes (number of features) equal to 50, 200, and 500, respectively in Table \ref{RESULTSACC2}. The experimental setup of \cite{NeurIPS2019} is exactly the same as ours. Computations were ran on Amazon Web Services (AWS) using two c5.9x-large EC2 instances with 36 vCPUs, 72.0 GiB of memory and 32 threads connected over a Gigabit ethernet local area network. These runtimes were computed for an average number of words in Alice's text equal to 21.7  words. We note that for 500 features, the best performing protocol proposed in \cite{NeurIPS2019} has a total runtime of 10.4s, this runtime is divided in 6.6s for feature extraction and 3.8s for the classification of the feature vector. Our proposed protocol has a runtime of 0.147s using a larger number of features (688) and a much larger number of words in Alice's text (160).}

\change{Since the protocols in \cite{NeurIPS2019} were implemented in Java and ours are implemented in Rust, one could wonder if the difference in performance is solely due to the change in the programming language. In order to check the effect of the programming language's choice, we have implemented the feature vector classification protocols proposed in \cite{NeurIPS2019} in Rust. The results are presented in Table \ref{RESULTSACC3}. We can observe that the resulting runtimes are faster than the Java ones. However, they are still 4 times (Logistic Regression) and 6 times (AdaBoost) slower than our proposed classification protocol.}

\begin{table*}
    \caption{Runtime for the secure text classification protocol proposed in \cite{NeurIPS2019} (Java framework), broken down in time needed for feature vector extraction (Extr) and time for feature vector classification (Class).}
    \centering
    \small{
    {\begin{tabular}{l|c r r r}
    \textbf{Java Implementation \cite{NeurIPS2019}} & \multicolumn{3}{c}{\textbf{Runtime (sec)}}\\
    \hline
    \hline
    & Extr & Class & Tot \\
    \hline
    Ada; 50 trees; depth 1   & 0.8 & 6.4 & 7.2\\ 
    Ada; 200 trees; depth 1    & 2.8 & 6.4 & 9.2\\
    Ada; 500 trees; depth 1  & 6.6 & 6.7 & 13.3\\
    \hline
    Logistic regression (50 feat.)   & 0.8 & 3.7 & 4.5\\
    Logistic regression (200 feat.)  & 2.8 & 3.7 & 6.5 \\
    Logistic regression (500 feat.)  & 6.6 & 3.8 & 10.4\\
    \end{tabular}
    }}
    \label{RESULTSACC2}
\end{table*}

\begin{table*}
    \caption{Runtime for the feature vector classification of a text from \cite{NeurIPS2019} implemented in Rust.}
    \centering
    \small{
    {\begin{tabular}{l| c r}
    \textbf{Rust Implementation} & \multicolumn{1}{c}{\textbf{Runtime (msec)}}\\
    \hline
    \hline
    & Class \\
    \hline
    Ada; 50 trees; depth 1   & 62\\ 
    Ada; 200 trees; depth 1    & 62\\
    Ada; 500 trees; depth 1  & 66\\
    \hline
    Logistic regression (50 feat.)   & 38\\
    Logistic regression (200 feat.)  & 39 \\
    Logistic regression (500 feat.)  & 41\\
    \end{tabular}}
    }
    \label{RESULTSACC3}
\end{table*}

As the protocols presented by Reich et al.~\cite{NeurIPS2019}, our solution does not leak any information about Alice’s words to Bob neither the words of Bob's model for Alice, and classifies an SMS as ham or spam (even for a model with 5200 features) in less than 0.3s, in the worst case, and less than 0.022s for an average message of our data set, while using the same type of machines that they used. Our results include communication and computation times. 

More recently, Badawi \emph{et. al} proposed a protocol for privacy-preserving text classification based on fully homomorphic encryption \cite{al2020privft}. They obtained a highly efficient, GPU-accelerated implementation that improves the state-of-the-art of FHE based inference by orders of magnitude. A GPU equipped machine can compute the private classification of a text message in about 0.17 seconds in their implementation. This time does not include the communication time to send the encrypted text from the client to the server and the time to receive the result. In a Gigabit Ethernet network, that would probably add anything between 0.3s to 0.5s to their total running time because of the ciphertext expansion resulting from the use of FHE.    

\section{Conclusion}

Privacy-preserving machine learning protocols are powerful solutions to perform operations on data while maintaining the privacy of the data. To the best of our knowledge, we propose the first privacy-preserving Naive Bayes classifier with private feature extraction. No information is revealed regarding either Bob's model (including which words belong to the model) or the words contained in Alice's SMS.
Our Rust implementation provides a fast and secure solution for the classification of unstructured text. Applying our solution to the case of spam detection, we can classify an SMS as spam or ham in less than 340\,ms in the case where the dictionary size of Bob's model includes all words ($n = 5200$) and Alice's SMS has at most $m = 160$ unigrams. In the case with $n = 369$ and $m = 8$ (the average of a spam SMS in the database), our solution takes only 21\,ms. Besides, the accuracy is practically the same as performing the Naive Bayes classification in the clear. It is important to note that our solution can be used in any application where Naive Bayes can be used. Thus, we believe that our solution is practical for the privacy-preserving classification of unstructured text. To the best of our knowledge, our solution is the fastest SMC based solution for private text classification. 

\change{Finally, we would like to pint out that whenever Alice is provided with the output of the classification, she will learn some information about Bob's model. This is unavoidable but does not contradict our security definition. Indeed, such feature is present in the ideal functionality used to define the security of our proposed classification protocol. A way to decrease such release of information would be to add differential privacy to the model, so that Alice would never be able to tell with certainty when a word belongs to Bob dictionary or not. That would decrease Alice's information about Bob's dictionary at the cost of reducing the accuracy of the model. We leave these questions as future work.}

\end{document}